\documentclass[twoside,leqno,twocolumn]{article}  
\usepackage{ltexpprt} 
\usepackage{amsmath}
\DeclareMathOperator*{\argmin}{arg\,min}
\renewcommand{\vec}[1]{\mathbf{#1}}
\usepackage{algorithm}
\usepackage{algorithmic}
\usepackage{multirow} 
\usepackage{graphicx}
\usepackage{subfigure}
\usepackage[flushleft]{threeparttable}
\usepackage[hyphens]{url}

\begin{document}

\title{\Large Top-N Recommendation with Novel Rank Approximation}
\author{Zhao Kang\thanks {Department of Computer Science, Southern Illinois University, Carbondale, IL 62901, USA. Zhao.Kang@siu.edu, qcheng@cs.siu.edu. Qiang Cheng is the corresponding author. } \and Qiang Cheng\footnotemark[1]}

\date{}

\maketitle


\begin{abstract} 
The importance of accurate recommender systems has been widely recognized by academia and industry. However, the recommendation quality is still rather low. Recently, a linear sparse and low-rank representation of the user-item matrix has been applied to produce Top-$N$ recommendations. This approach uses the nuclear norm as a convex relaxation for the rank function and has achieved better recommendation accuracy than the state-of-the-art methods. In the past several years, solving rank minimization problems by leveraging nonconvex relaxations has received increasing attention. Some empirical results demonstrate that it can provide a better approximation to original problems than convex relaxation. In this paper, we propose a novel rank approximation to enhance the performance of Top-$N$ recommendation systems, where the approximation error is controllable. Experimental results on real data show that the proposed rank approximation improves the Top-$N$ recommendation accuracy substantially.
\end{abstract}

\section{Introduction}
 Learning about users' preference and making recommendations for them is of great importance in e-commerce, targeted advertising and web search. Recommendation is rapidly becoming one of the most successful applications of data mining and machine learning. The goal of a Top-$N$ recommendation algorithm is to produce a length-$N$ list of recommended items such as movies, music, and so on. Over the years, a number of algorithms have been developed to tackle the Top-$N$ recommendation problem \cite{ricci2011introduction}. They make predictions based on the user feedback, for example, purchase, rating, review, click, check-in, etc. The existing methods can be broadly classified into two classes: content-based filtering \cite{balabanovic1997fab} and collaborating filtering (CF) \cite{gu2010collaborative} \cite{wang2006recommendation} \cite{zhang2006learning}.

Content-based filtering: in this approach, features or descriptions are utilized to describe the items and a user profile or model is built using the past item rating to summarize the types of items this user likes \cite{pazzani2007content}. This approach is based on an underlying assumption that liking a feature in the past leads to liking the feature in the future. Some disadvantages of this approach are: If the content does not contain enough information to discriminate the items, then the recommendation will not be accurate; When there is not enough information to build a solid model for a new user, the recommendation will also be jeopardized. 

Collaborating filtering: in this approach, user/item co-rating information is utilized to build models. Specifically, CF relies on the following  assumption: if a user A likes some items that are also liked by another user B, A is likely to share the same preference with B on another item \cite{desrosiers2011comprehensive}. One challenge for CF algorithms is to have the ability to deal with highly sparse data, since users typically rate only a small portion of the available items. 

In general, CF methods can be further divided into two categories: nearest-neighborhood-based methods and model-based methods. The first class of methods compute the similarities between the users/items using the co-rating information and new items are recommended based on these similarities \cite{ning2011slim}. One representative method of this kind is Item-based k-nearest-neighbor (ItemKNN) \cite{deshpande2004item}. On the other hand, model-based methods employ a machine learning algorithm to build a model, which is then used to perform the recommendation task \cite{kang2016top}. This model learns the similarities between items or latent factors that explain ratings. For example, matrix factorization (MF) method uncovers a low-rank latent structure of data, approximating the user-item matrix as a product of two factor matrices. 

Matrix factorization is popular for collaborative prediction and many works are based on it. For instance, pure singular-value-decomposition-based (PureSVD) \cite{cremonesi2010performance} MF method represents users and items by the most principal singular vectors of the user-item matrix; weighted regularized matrix factorization (WRMF) \cite{pan2008one} method deploys a weighting matrix to discriminate between the contributions from observed purchase/rating activities and unobserved ones. 

Recently, a novel Top-$N$ recommendation method has been developed, called LorSLIM \cite{cheng2014lorslim}, which has been shown to achieve good performance on a wide variety of datasets and outperform other state-of-the-art approaches. LorSLIM improves upon the traditional item-based nearest neighbor CF approaches by learning directly from the data, a sparse and low-rank matrix of aggregation coefficients that are analogous to the traditional item-item similarities. It demonstrates that low-rank requirement on the similarity matrix is crucial to improve recommendation quality. Since the rank function can hardly be used directly, the nuclear norm \cite{recht2010guaranteed} is adopted as a convex relaxation of  the matrix rank function in LorSLIM. Although the nuclear norm indeed recovers low-rank matrices in some scenarios \cite{candes2009exact}, some recent work has pointed out that this relaxation may lead to poor solutions \cite{shi2011limitations} \cite{kang2015pca} \cite{kang2015cikm} \cite{srebro2010collaborative}. In this paper, we propose a novel relaxation which provides a better approximation to the rank function than the nuclear norm. By using this new approximation in LorSLIM model, we observe significant improvement over the current methods. The main contributions of
our paper are as follows:
\begin{enumerate}
\item We introduce a novel matrix rank approximation function, whose value can be very close to the real rank. This can be applied in a range of rank minimization problems in machine learning and computer vision.
\item An efficient optimization strategy is designed for this associated nonconvex optimization problem, which admits a closed-form solution to every subproblem. 
\item As an illustration, we perform experiments on six real datasets. It indicates that our Top-$N$ recommendation approach considerably outperforms the state-of-the-art algorithms which give similar performances on most datesets. Thus this fundamental enhancement is due to our better rank approximation.   
\end{enumerate}
The remainder of this paper is organized as follows. In Section \ref{notation}, we give some notations. Section \ref{related} describes related work. Section \ref{proposed} introduces the proposed model. In Section \ref{expeval}, we describe our experimental framework. Experimental results and analysis are presented in Section \ref{discuss}; Section \ref{conclude} draws conclusions. 

\section{Notations and Definitions}
\label{notation}
Let $U=\{u_1,u_2,...,u_m\}$ and $T=\{t_1,t_2,...,t_n\}$ represent the sets of all users and all items, respectively. The entire set of user-item purchases/ratings is to be represented by user-item matrix $X$ of size $m\times n$. The value of $x_{ij}$ is 1 or a positive value if user $u_i$ has ever purchased/rated item $t_j$; otherwise it is $0$. $\vec{x}_i^T$ , the $i$-th row of $X$, denotes the purchase/rating history of user $u_i$ on all items. The $j$-th column of $X$ denoted as $\vec{x}_j$ is the purchase/rating history of all users on item $t_j$. The aggregation coefficient matrix is represented as $W$ of size $n\times n$. $\vec{w}_j$ is a size-$n$ column vector of aggregation coefficients. $\|W\|_1=\sum\limits_{i}\sum\limits_j|w_{ij}|$ is the $l_1$-norm of $W$. $\|W\|_F^2=\sum\limits_{i}\sum\limits_j w_{ij}^2$ denotes the squared Frobenius norm of $W$. The nuclear norm of $W$ is $\|W\|_*=\sum\limits_{i=1}^n\sigma_i(W)$, where $\sigma_i$ is the $i$-th singular value of $W$. The unit step function $s(x)$ has value 1 for $x>0$ and 0 if $x=0$. The rank of matrix $W$ is $\sum\limits_{i=1}^n s(\sigma_i(W))$. We use $\sigma(W)$ to denote the vector of all singular values of $W$ in non-increasing order. Moreover, $I$ denotes the identity matrix.

In this paper, we denote all vectors (e.g., $\vec{x}_i$, $\vec{x}_j$) with bold lower case letters. We represent all matrices (e.g. $W$, $X$) with upper case letters. A predicted value is represented by having a $\wedge$ mark.
\section{Relevant Research}
\label{related}
Recently, an interesting Top-$N$ recommendation method, sparse linear methods (SLIM) has been proposed \cite{ning2011slim} which generates recommendation lists by learning a sparse similarity matrix. SLIM solves the following regularized optimization problem: 
\begin{equation}
\begin{split}
&\min_W \frac{1}{2}\|X-XW\|_F^2+\frac{\beta}{2}\|W\|_F^2+\lambda \|W\|_1  \\
&s.t.\quad W\ge 0, \quad diag(W)=0,   
\end{split}
\label{slim}
\end{equation}
where the first term measures the reconstruction error, $\|W\|_1$ enforces the sparsity on $W$, and the second and third terms combine the sparsity-inducing property of $\|W\|_1$ with the smoothness of $\|W\|_F^2$, in a way similar to the elastic net \cite{zou2005regularization}. The first constraint is intended to ensure that the learned coefficients represent positive similarities between items, while the second constraint is applied to avoid the trivial solution in which $W$ is an identity matrix, i.e., an item always recommends itself. It has been shown that SLIM outperforms other Top-$N$ recommendation methods. A drawback of SLIM is that it can only model relations between items that have been co-purchased/co-rated by at least one user \cite{cheng2014lorslim}. Therefore, it fails to capture the potential dependencies between items that have not been co-rated by at least one user, while modeling relations between items that are not co-rated is essential for good performance of item-based approaches in sparse datasets. 

To address the above issue, LorSLIM \cite{cheng2014lorslim} further considers the low-rank structure of $W$. This idea is inspired by the factor model, which assumes that a few latent variables are responsible for items' features $F$ and the coefficient matrix factors, $W\approx FF^T$, with $F$ being of low-rank. Finally, together with sparsity, it constructs a block diagonal $W$, i.e., the items have been classified into many smaller "clusters" or categories. This situation happens frequently in real life such as movies, music, books and so on. Therefore, this model promotes the recommendation precision further. 

In LorSLIM, the nuclear norm $\|W\|_*$ is utilized as a surrogate for the rank of $W$. By comparing $\sum\limits_{i=1}^n\sigma_i$ with $\sum\limits_{i=1}^n s(\sigma_i)$, we can see that when the singular values are much larger than 1, the nuclear norm approximation deviates from the true rank markedly. The nuclear norm is essentially an $l_1$-norm of the singular values and it is well known that $l_1$-norm has a shrinkage effect and leads to a biased estimator \cite{fan2001variable} \cite{zhang2010nearly}. Recently, some variations of the nuclear norm have been studied, e.g., some of the largest singular values are subtracted from the nuclear norm in truncated nuclear norm regularization \cite{hu2013fast}; a soft thresholding
rule is applied to all singular values in singular value thresholding algorithm \cite{cai2010singular}; some generalized nonconvex rank approximations have been investigated in \cite{lu2014generalized} \cite{lu2015generalized}. In some applications, they show good performance; however, these models are either overly simple or only restricted to some specific applications.

In this paper, we develop a more general approach, which directly approximates the rank function with our formulation and optimization. Then we show that better rank approximation can improve the recommendation accuracy substantially.
\section{Proposed Framework}
\label{proposed}
\subsection{Problem Setup}
In this paper, we propose the following continuous function to replace the unit step function $s(x)$ in the definition of the rank function:
\begin{equation}
\label{app}
 f(x)=\sum_i (1-\mathrm{e}^{-|x_i|/\delta}), 
\end{equation}
where $\delta>0$ controls the approximation accuracy. Equation (\ref{app}) is similar to the formulation proposed in \cite{malek2014iterative}. For any $\sigma_i\in [0, \infty)$, $1-\mathrm{e}^{-\sigma_i/\delta}\in [0, 1)$, and $\int_0^\infty |(1-\mathrm{e}^{-\sigma_i/\delta})-s(\sigma_i)|^2\,d\sigma_i=\frac{\delta}{2}$; hence, for any matrix $X$,  $f(\sigma(X))$ approaches its true rank as $\delta$ approaches zero. 

There are several motivations behind this formulation. First, 
 it attenuates the contributions from large singular values significantly, thus overcomes the imbalanced penalization of different singular values.  
 Second, by defining $\frac{\partial f(0)}{\partial x_i}=\lim\limits_{x_i\to 0^+}\frac{1}{\delta}\mathrm{e}^{-x_i/\delta}=\frac{1}{\delta}$, $f$ is differentiable and concave in $[0, \infty)$. Third, 
$F(X)=f(\sigma(X))$ is unitarily invariant. The last two properties facilitate subsequent optimization and computation much. Compared to many other approaches \cite{kang2015robust} \cite{peng2015subspace} \cite{lu2014generalized}, this formulation enjoys simplicity and efficacy.

To alleviate the issues associated with the nuclear norm, we solve the following problem for Top-$N$ recommendation task:
\begin{equation}
\begin{split}
&\min_W \frac{1}{2}\|X-XW\|_F^2+\alpha \|W\|_1 +\beta \sum\limits_{i=1}^n (1-e^{-\sigma_i(W)/\delta})\\
&+l_{\mathcal{R}_+}(W) \\
&s.t.\quad diag(W)=0.   
\end{split}
\label{ourmodel}
\end{equation}
Different from \cite{ning2011slim} \cite{cheng2014lorslim}, we incorporate the nonnegative constraint into the objective function by making use of $l_{\mathcal{R}_+}$, which is defined element-wisely as
\begin{eqnarray*}
\label{error21}
l_{\mathcal{R}_+}(x)=\left\{
\begin{array}{ll} 0, & \mbox{if $x\geq 0$};\\
+\infty, & \mbox{otherwise.}
\end{array}\right.
\end{eqnarray*}
\subsection{Optimization}
Since (\ref{ourmodel}) is a nonconvex problem, it is hard to solve directly. We introduce auxiliary variables to make the objective function separable and solve the following equivalent problem:
\begin{equation}
\begin{split}
&\min_W \frac{1}{2}\|X-XW\|_F^2+\alpha \|Z_1\|_1 +\beta \sum\limits_{i=1}^n (1-e^{-\sigma_i(Z_2)/\delta})\\
&+l_{\mathcal{R}_+}(Z_3) \\
&s.t.\quad diag(W)=0,\hspace{.05cm} W=Z_1,\hspace{.05cm} W=Z_2,\hspace{.05cm} W=Z_3.  
\end{split}
\label{newmodel}
\end{equation}
This can be solved by using the augmented Lagrange multiplier (ALM) method \cite{bertsekas1999nonlinear}. We turn to minimizing the following
augmented Lagrangian function:
\begin{equation}
\begin{aligned}
\lefteqn{\mathcal{L}(W, Z_1, Z_2, Z_3)}\\
&=\frac{1}{2}\|X-XW\|_F^2+\alpha \|Z_1\|_1+\beta\sum\limits_{i=1}^n (1-e^{-\sigma_i(Z_2)/\delta})\\
&+l_{\mathcal{R}_+}(Z_3)+\frac{\mu}{2}\Big[\|Z_1-(W-\frac{Y_1}{\mu})\|_F^2+\\
&\|Z_2-(W-\frac{Y_2}{\mu})\|_F^2+\|Z_3-(W-\frac{Y_3}{\mu})\|_F^2\Big],
\end{aligned}
\label{lag}
\end{equation}
where $\mu> 0$ is the penalty parameter and $Y_1$, $Y_2$, $Y_3$ are the Lagrange multipliers. This unconstrained problem can be minimized with respect
to $Z_1$, $Z_2$, and $Z_3$ alternatively, by fixing the other variables, and then updating the Lagrange multipliers $Y_1$, $Y_2$, and $Y_3$. At the $(t + 1)$th iteration,
\begin{equation}
\begin{split}
&W^{t+1}=\argmin_W \frac{1}{2}\|X-XW\|_F^2+\frac{\mu^t}{2}\Big[\|Z_1^t-(W-\frac{Y_1^t}{\mu})\|_F^2\\
&+\|Z_2^t-(W-\frac{Y_2^t}{\mu})\|_F^2+\|Z_3^t-(W-\frac{Y_3^t}{\mu})\|_F^2\Big]
\end{split}
\label{wmini}
\end{equation}
We can see that the objective function of (\ref{wmini}) is quadratic and strongly convex in $W$, which has a closed-form solution:
\begin{equation}
\label{solw}
\begin{split}
&W^{t+1}=(3\mu^t I+X^TX)^{-1}[\mu^t(Z_1^t+Z_2^t+Z_3^t)+\\
&(Y_1^t+Y_2^t+Y_3^t)+X^TX].
\end{split}
\end{equation}
For $Z_1$ minimization, we have
\begin{equation}
Z_1^{t+1}=\argmin_{Z_1} \alpha \|Z_1\|_1+\frac{\mu^t}{2}\left\|Z_1-(W^{t+1}-\frac{Y_1^t}{\mu})\right\|_F^2,
\end{equation}
which can be solved by the following lemma \cite{beck2009fast}.
\begin{lemma}
For $\mu>0$ and $Y\in\mathcal{R}^{m\times n}$, the solution
of the problem
\[
\min_X \mu\|X\|_1+\frac{1}{2}\|X-Y\|_F^2\]
is given by $X_\mu(Y)$, which is defined component-wisely by
\[
[X_\mu(Y)]_{ij}= \textrm{max}\{|y_{ij}|-\mu, 0\} \cdot sign(y_{ij}).\]
\end{lemma}
Therefore, by letting $Q=W^{t+1}-\frac{Y_1^t}{\mu^t}$, we can solve $Z_1$ element-wisely as below:
\begin{equation}
\label{solz1}
(Z_1^{t+1})_{ij}=\textrm{max}(|Q_{ij}|-\alpha/\mu^t,0)\cdot sign(Q_{ij}).
\end{equation}
To update $Z_2$, we have 
\begin{equation}
\min_{Z_2} \beta \sum\limits_{i=1}^n (1-e^{-\sigma_i(Z_2)/\delta})+\frac{\mu^t}{2}\|Z_2-(W^{t+1}-\frac{Y_2^t}{\mu^t})\|_F^2, 
\label{solz2}
\end{equation} 
This can be solved with the following theorem.
\begin{theorem}
\cite{kang2015logdet}
\label{firsthm}
 If $F(Z)=f(\sigma(Z))$ is a unitarily invariant function, $\mu>0$, and $A\in \mathbf{\mathcal{R}}^{m\times n}$ whose SVD is $U\Sigma_A V^T$ and $\Sigma_A=diag(\sigma_A)$ 
, then the optimal solution to the following problem
\begin{equation}
\min_Z F(Z)+\frac{\mu}{2}\left\|Z-A\right\|_F^2,
\label{theoremprob}
\end{equation}
is $Z^*$ with SVD being $U\Sigma_Z^*V^T$, where $\Sigma_Z^*=diag(\sigma^*)$ is obtained through the Moreau-Yosida operator $\sigma^* = \mathrm{prox}_{f, \mu} (\sigma_{A})$, defined as 
\begin{equation}
\label{scalar}
\mathrm{prox}_{f, \mu} (\sigma_A) := \argmin_{\sigma\geq0} f(\sigma) + \frac{\mu}{2}\|\sigma - \sigma_A\|_2^2.
\end{equation}
\end{theorem}
In our case, the first term in (\ref{scalar}) is concave while the second term is convex in $\sigma$, so we can resort to the difference of convex (DC) \cite{horst1999dc} optimization strategy. A linear approximation is applied at each iteration of DC programing. For this inner loop, at the $(k + 1)$th iteration,
\begin{equation}
\sigma^{k+1}=(\sigma_A-\frac{\beta\omega_k}{\mu^t})_+,
\end{equation}
where $\omega_k=\partial f(\sigma^k)$ is the gradient of $f(\cdot)$ at $\sigma^k$ and $U diag\lbrace\sigma_A\rbrace V^T$ is the SVD of $W^{t+1}-\frac{Y_2^t}{\mu^t}$. Finally, it converges to a local optimal point $\sigma^*$. Then $Z_2^{t+1}=U diag\lbrace\sigma^*\rbrace V^T$.

To update $Z_3$, we need to solve 
\begin{equation}
Z_3^{t+1}=\argmin_{Z_3} l_{\mathcal{R}_+}(Z_3)+\frac{\mu^t}{2}\|Z_3-(W^{t+1}-\frac{Y_3^t}{\mu^t})\|_F^2,
\end{equation} 
which yields the updating rule 
\begin{equation}
\label{solz3}
Z_3^{t+1}=\textrm{max}(W^{t+1}-\frac{Y_3^t}{\mu^t}, 0).
\end{equation}
Here max$(\cdot)$ is an element-wise operator. The complete procedure
is outlined in Algorithm 1.
\begin{algorithm}[tb]
   \caption{Solve (\ref{ourmodel})}
   \label{alg:rankminimization}
  {\bfseries Input:} Original data matrix $X\in \mathbf{\mathcal{R}}^{m\times n}$, parameters $\alpha>0$, $\beta>0$, $\mu^0>0$, $\gamma>1$.\\
{\bfseries Initialize:} $Z_1=Z_2=Z_3$ as $n$-by-$n$ matrices with random numbers between 0 and 1, $Y_1=Y_2=Y_3=0$.\\
  {\bfseries REPEAT}
\begin{algorithmic}[1]
   \STATE Obtain $W$ through (\ref{solw}).
   \STATE Update $Z_1$ as (\ref{solz1}).
  \STATE Solve $Z_2$ by solving (\ref{solz2}).
   \STATE Update $Z_3$ as (\ref{solz3}).
\STATE Update the Lagrangian multipliers:
\begin{align*}
Y_1^{t+1}&=Y_1^{t}+\mu^t(Z_1^{t+1}-W^{t+1}),\\
Y_2^{t+1}&=Y_2^{t}+\mu^t(Z_2^{t+1}-W^{t+1}),\\
Y_3^{t+1}&=Y_3^{t}+\mu^t(Z_3^{t+1}-W^{t+1}).
\end{align*}
\STATE Update the parameter $\mu^t$ by $\mu^{t+1}=\gamma\mu^t$.
\end{algorithmic}
\textbf{ UNTIL} {stopping criterion is met.}
\end{algorithm}
\section{Experimental Evaluation}
\label{expeval}
\subsection{Datasets}
\begin{table}[ht]
\caption{The datasets used in evaluation}
\label{tab:data}
\begin{center}
\resizebox{.45\textwidth}{!}{
\begin{tabular}{llllllll}
\multicolumn{1}{c}{dataset}  &\multicolumn{1}{c}{\#users} &\multicolumn{1}{c}{ \#items}  &\multicolumn{1}{c}{\#trns} &\multicolumn{1}{c}{rsize}  &\multicolumn{1}{c}{csize}&\multicolumn{1}{c}{ density}&\multicolumn{1}{c}{ratings}\\
\hline\hline \\
Delicious&1300&4516&17550&13.50&3.89&0.29\%&-\\
 lastfm&8813&6038&332486&37.7&55.07&0.62\%&- \\
BX&4186&7733&182057&43.49&23.54&0.56\%&- \\
\hline \\
ML100K &943&1682&100000&106.04&59.45&6.30\%&1-10\\
Netflix&6769&7026&116537&17.21&16.59&0.24\%&1-5 \\
Yahoo&7635&5252&212772&27.87&40.51&0.53\% &1-5 \\
\hline
\end{tabular}}
   \begin{tablenotes}
      \small
      \item The ``\#users", ``\#items", ``\#trns" columns show the number of users, number of items and number of transactions, respectively, in each dataset. The ``rsize" and ``csize" columns show the average number of ratings of each user and of each item, respectively, in each dataset. Column corresponding to ``density" shows the density of each dataset (i.e., density=\#trns/(\#users$\times$\#items)). The ``ratings" column is the rating range of each dataset with granularity 1. 
    \end{tablenotes}
\end{center}
\end{table}
We evaluate the performance of our method on six different real datasets whose characteristics are summarized in Table \ref{tab:data}. These datasets represent different applications of a recommendation algorithm. They can be broadly categorized into two classes.

The first class contains Delicious, lastfm and BX. These three datasets have only implicit feedback, i.e., they are represented by binary matrices. Specifically, Delicious was the bookmarking and tagging information of 2$K$ users in Delicious social bookmarking system\footnote{\url{http://www.delicious.com}}, in which each URL was bookmarked by at least 3 users. Lastfm represents music artist listening information extracted from the last.fm online music system\footnote{\url{ http://www.last.fm }}, in which each music artist was listened to by at least 10 users and each user listened to at least 5 artists. BX is a part of the Book-Crossing dataset\footnote{\url{http://www.informatik.uni-freiburg.de/~cziegler/BX/}} such that only implicit interactions were contained and each book was read by at least 10 users. 

The second class contains ML100K, Netflix and Yahoo. All these datasets contain multi-value ratings. Specifically, the ML100K dataset contains movie ratings and is a subset of the MovieLens research project\footnote{\url{http://grouplens.org/datasets/movielens/}}. The Netflix is a subset of  Netflix Prize dataset\footnote{\url{http://www.netflixprize.com/}} and each user rated at least 10 movies. The Yahoo dataset is a subset obtained from Yahoo!Movies user ratings\footnote{\url{http://webscope.sandbox.yahoo.com/catalog.php?datatype=r}}. In this dataset, each user rated at least 5 movies and each movie was rated by at least 3 users.

\begin{table*}[!ht]
\begin{center}
\begin{threeparttable}
\caption{Comparison of Top-N recommendation algorithms}
\label{tab:comp}
\small 
\begin{tabular}{llllllllllllll}
\hline
\multirow{2}{*}{method} &
\multicolumn{6}{c}{Delicious} &
\multicolumn{1}{c}{}&
\multicolumn{6}{c}{lastfm} \\
\cline{2-7} \cline{9-14} 
  & \multicolumn{4}{c}{params} &\multicolumn{1}{c}{HR}  & \multicolumn{1}{c}{ARHR} &\multicolumn{1}{c}{}&\multicolumn{4}{c}{params} &\multicolumn{1}{c}{HR}  & \multicolumn{1}{c}{ARHR} \\ 
\hline
ItemKNN&300&-&-&-&0.300&0.179&&100&-&-&-&0.125&0.075\\
PureSVD&1000&10&-&-&0.285&0.172&&200&10&-&-&0.134&0.078\\
WRMF&250&5&-&-&0.330&0.198&&100&3&-&-&0.138&0.078\\
BPRKNN&1e-4&0.01&-&-&0.326&0.187&&1e-4&0.01&-&-&0.145&0.083\\
BPRMF&300&0.1&-&-&0.335&0.183&&100&0.1&-&-&0.129&0.073\\
SLIM&10&1&-&-&0.343&0.213&&5&0.5&-&-&0.141&0.082\\
LorSLIM&10&1&3&3&0.360&0.227&&5&1&3&3&0.187&0.105\\
Our&20&5&20&-&\bf{0.385}&\bf{0.232}&&10&0.1&10&-&\bf{0.210}&\bf{0.123}\\
\hline\hline
\multirow{2}{*}{method} &
\multicolumn{6}{c}{BX} &
\multicolumn{1}{c}{}&
\multicolumn{6}{c}{ML100K} \\
\cline{2-7} \cline{9-14} 
  & \multicolumn{4}{c}{params} &\multicolumn{1}{c}{HR}  & \multicolumn{1}{c}{ARHR} &\multicolumn{1}{c}{}&\multicolumn{4}{c}{params} &\multicolumn{1}{c}{HR}  & \multicolumn{1}{c}{ARHR} \\ 
\hline
ItemKNN&400&-&-&-&0.045&0.026&&10&-&-&-&0.287&0.124\\
PureSVD&3000&10&-&-&0.043&0.023&&100&10&-&-&0.324&0.132\\
WRMF&400&5&-&-&0.047&0.027&&50&1&-&-&0.327&0.133\\
BPRKNN&1e-3&0.01&-&-&0.047&0.028&&2e-4&1e-4&-&-&0.359&0.150\\
BPRMF&400&0.1&-&-&0.048&0.027&&200&0.1&-&-&0.330&0.135\\
SLIM&20&0.5&-&-&0.050&0.029&&2&2&-&-&0.343&0.147\\
LorSLIM&50&0.5&2&3&0.052&0.031&&10&8&5&3&0.397&0.207\\
Our&1&1&10&-&\bf{0.061}&\bf{0.038}&&200&0.2&700&-&\bf{0.434}&\bf{0.224}\\
\hline\hline
\multirow{2}{*}{method} &
\multicolumn{6}{c}{Netflix} &
\multicolumn{1}{c}{}&
\multicolumn{6}{c}{Yahoo} \\
\cline{2-7} \cline{9-14} 
  & \multicolumn{4}{c}{params} &\multicolumn{1}{c}{HR}  & \multicolumn{1}{c}{ARHR} &\multicolumn{1}{c}{}&\multicolumn{4}{c}{params} &\multicolumn{1}{c}{HR}  & \multicolumn{1}{c}{ARHR} \\ 
\hline
ItemKNN&200&-&-&-&0.156&0.085&&300&-&-&-&0.318&0.185\\
PureSVD&500&10&-&-&0.158&0.089&&2000&10&-&-&0.210&0.118\\
WRMF&300&5&-&-&0.172&0.095&&100&4&-&-&0.250&0.128\\
BPRKNN&2e-3&0.01&-&-&0.165&0.090&&0.02&1e-3&-&-&0.310&0.182\\
BPRMF&300&0.1&-&-&0.140&0.072&&300&0.1&-&-&0.308&0.180\\
SLIM&5&1.0&-&-&0.173&0.098&&10&1&-&-&0.320&0.187\\
LorSLIM&10&3&5&3&0.196&0.111&&10&1&2&3&0.334&0.191\\
Our&200&100&200&-&\bf{0.228}&\bf{0.122}&&300&10&100&-&\bf{0.360}&\bf{0.205}\\
\hline
\end{tabular}

   \begin{tablenotes}
      \small

      \item The parameters for each method are described as follows: ItemKNN: the number of neighbors $k$; PureSVD: the number of singular values and the number of SVD; WRMF: the dimension of the latent space and its weight on purchases; BPRKNN: its learning rate and regularization parameter $\lambda$; BPRMF: the latent space's dimension and learning rate; SLIM: the $l_2$-norm regularization parameter $\beta$ and the $l_1$-norm regularization coefficient
$\lambda$; LorSLIM: the $l_2$-norm regularization parameter $\beta$, the $l_1$-norm regularization parameter $\lambda$, the nuclear norm regularization coefficient $z$ and the auxiliary parameter $\rho$. Our: the $l_1$-norm regularization parameter $\alpha$, the rank regularization parameter $\beta$ and the auxiliary parameter $\mu^0$. $N$ in this table is 10. Bold numbers are the best performance in terms of HR and ARHR for each dataset. 

 \end{tablenotes}
\end{threeparttable}
\end{center}  
\end{table*}
\subsection{Evaluation Methodology}

To examine the effectiveness of the proposed method, we follow the procedure in \cite{ning2011slim} and adopt 5-fold cross validation. For each fold, a dataset is split into training and test sets by randomly selecting one non-zero entry for each user and putting it in the test set, while using the rest of the data for training the model\footnote{We use the same data as in \cite{cheng2014lorslim}, with partitioned datasets kindly provided by its first author.}. Then a ranked list of size-$N$ items for each user is produced. We then evaluate the model by comparing the ranked list of recommended items with the item in the test set. In the following results presented in this paper, $N$ is equal to 10 by default.

The recommendation quality is evaluated by the Hit Rate (HR) and the Average Reciprocal Hit Rank (ARHR) \cite{deshpande2004item}. HR is defined as
\begin{equation}
HR=\frac{\#\textrm{hits}}{\#\textrm{users}},
\end{equation}
where \#hits is the number of users whose item in the testing set is contained (i.e., hit) in the size-$N$ recommendation list, and \#users is the total number of users. An HR value of 1.0 means that the algorithm is able to always recommend hidden items correctly, whereas an HR value of 0.0 indicates that the algorithm is not able to recommend any of the hidden items. 

A drawback of HR is that it treats all hits equally without considering where they appear in the Top-$N$ list. ARHR addresses this by rewarding each hit based on its place in the Top-$N$ list, which is defined as:
\begin{equation}
ARHR=\frac{1}{\#\textrm{users}}\sum_{i=1}^{\#\textrm{hits}}\frac{1}{p_i},
\end{equation}
where $p_i$ is the position of the item in the ranked Top-$N$ list for the $i$-th hit. In this metric, hits that occur earlier in the ranked list are weighted higher than those occur later, and thus ARHR indicates how strongly an item is recommended. The highest value of ARHR is equal to HR which occurs when all the hits occur in the first position, and the lowest value is equal to HR/$N$ when all the hits occur in the last position of the list.

HR and ARHR are recommended as evaluation metrics since they directly measure the performance based on the ground truth data, i.e., what users have already provided feedback \cite{ning2011slim}. 

\begin{figure*}[!ht]
\centering
\subfigure[Delicious]{\includegraphics[width=.38\textwidth]{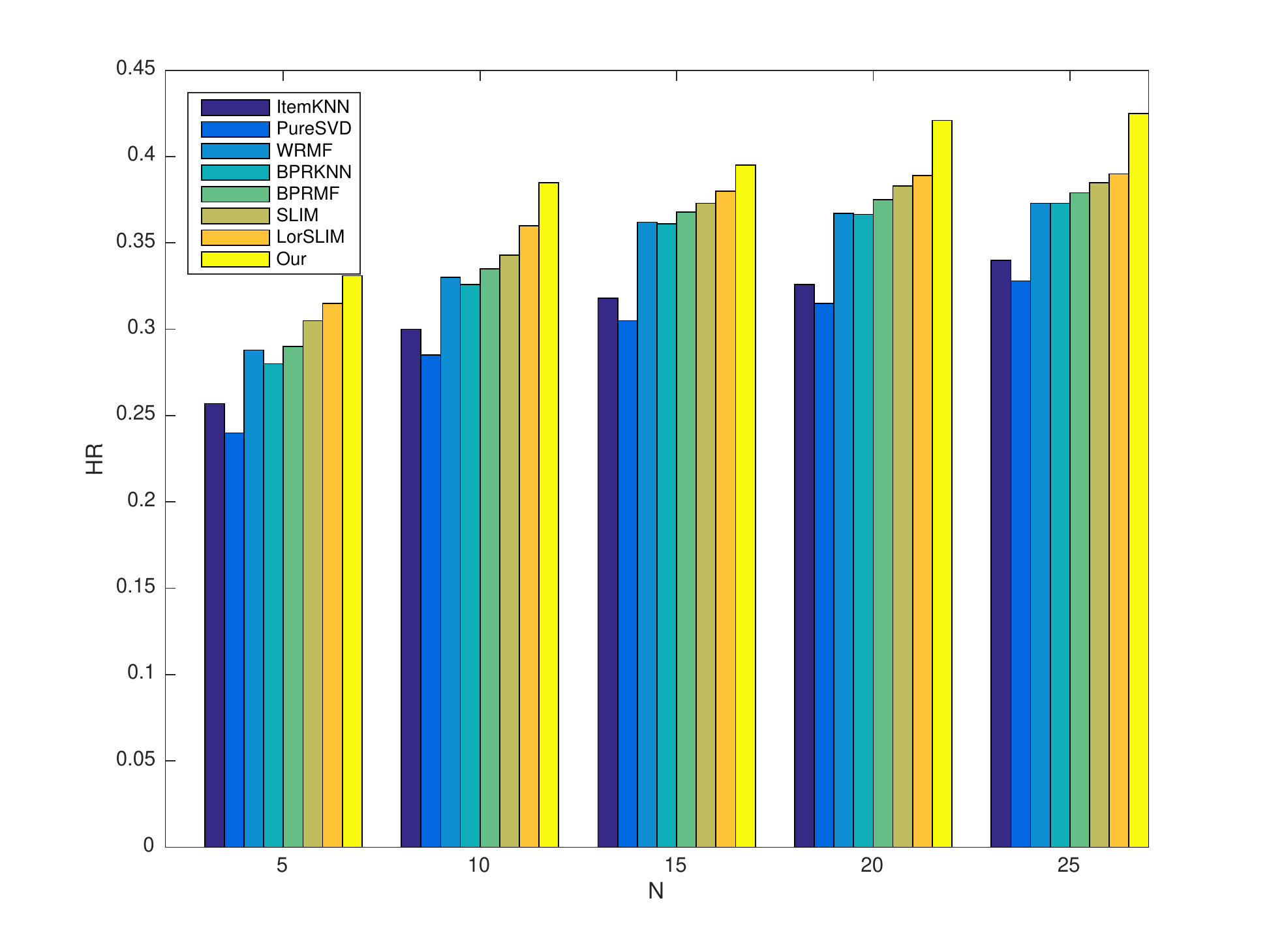}}
\subfigure[lastfm]{\includegraphics[width=.38\textwidth]{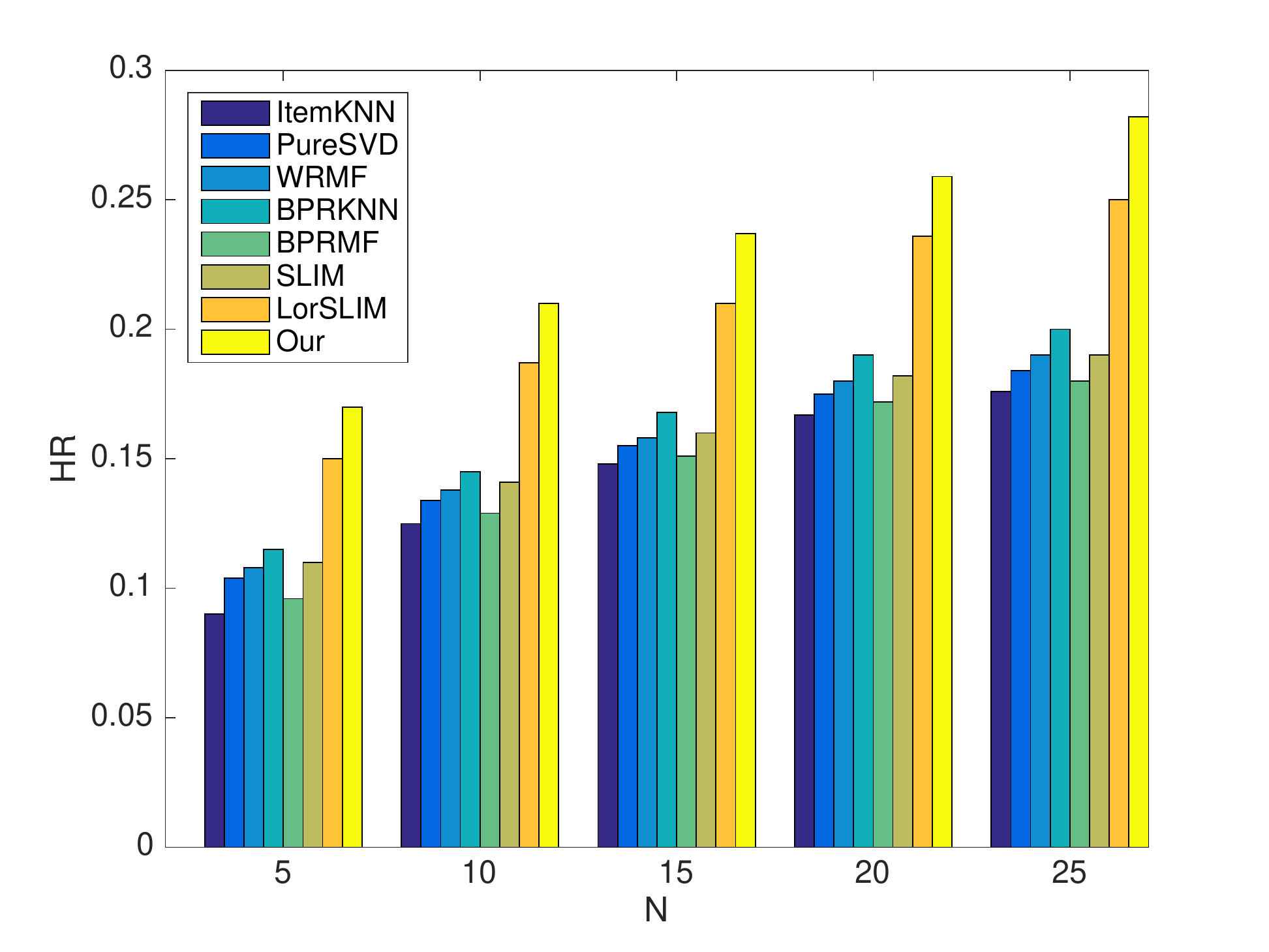}}\\
\subfigure[BX]{\includegraphics[width=.38\textwidth]{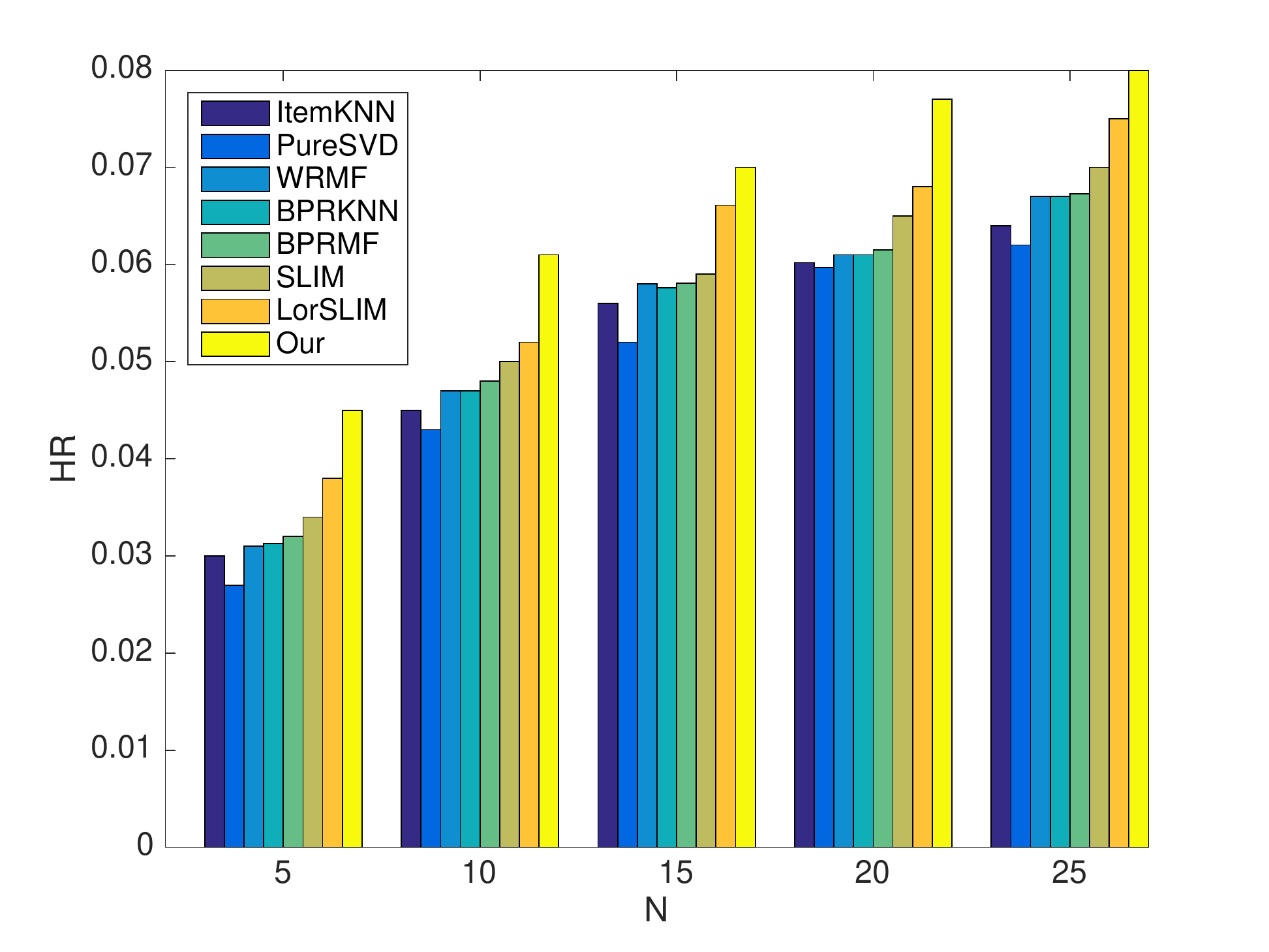}}
\subfigure[ML100K]{\includegraphics[width=.38\textwidth]{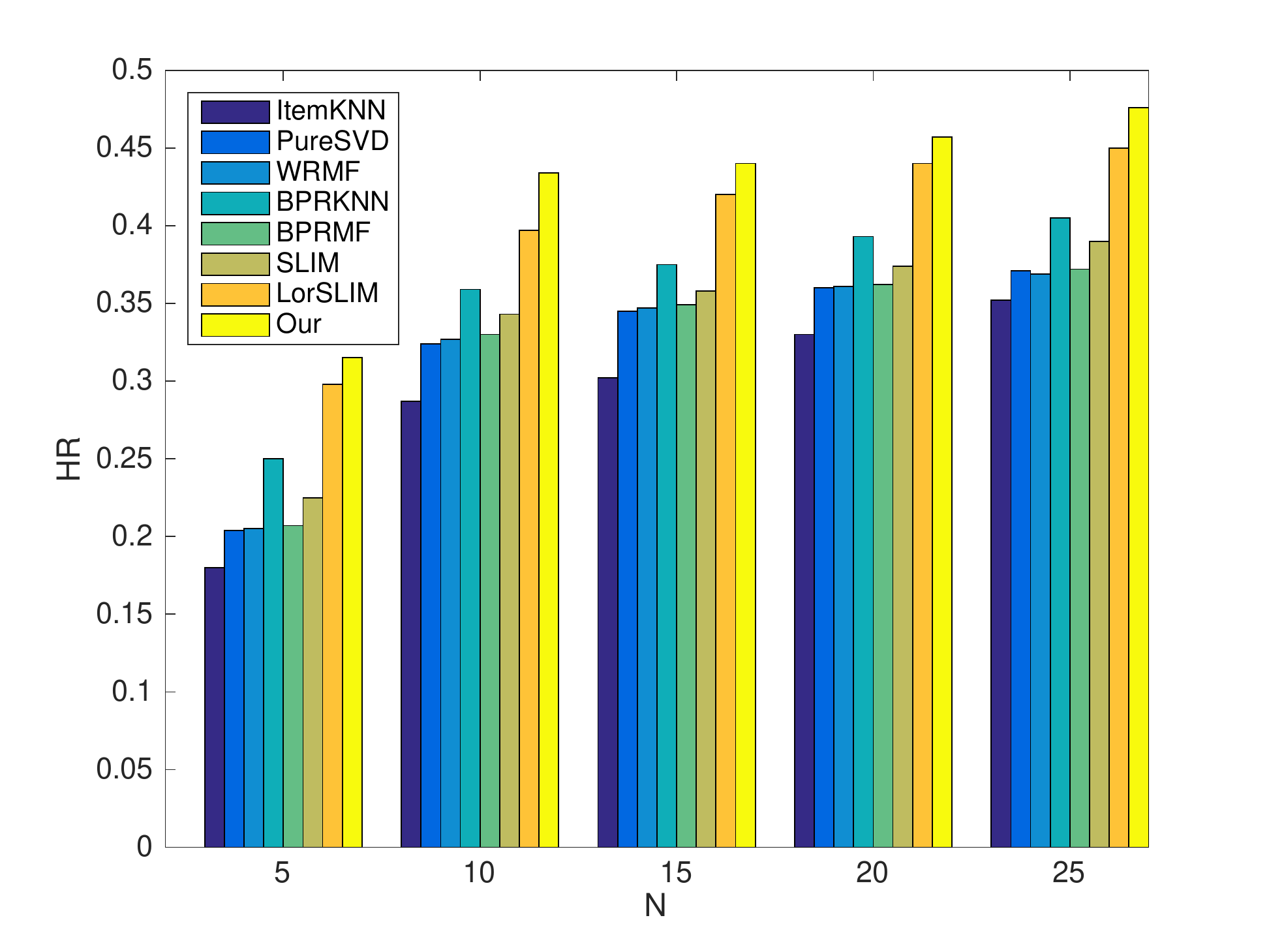}}\\
\subfigure[Netflix]{\includegraphics[width=.38\textwidth]{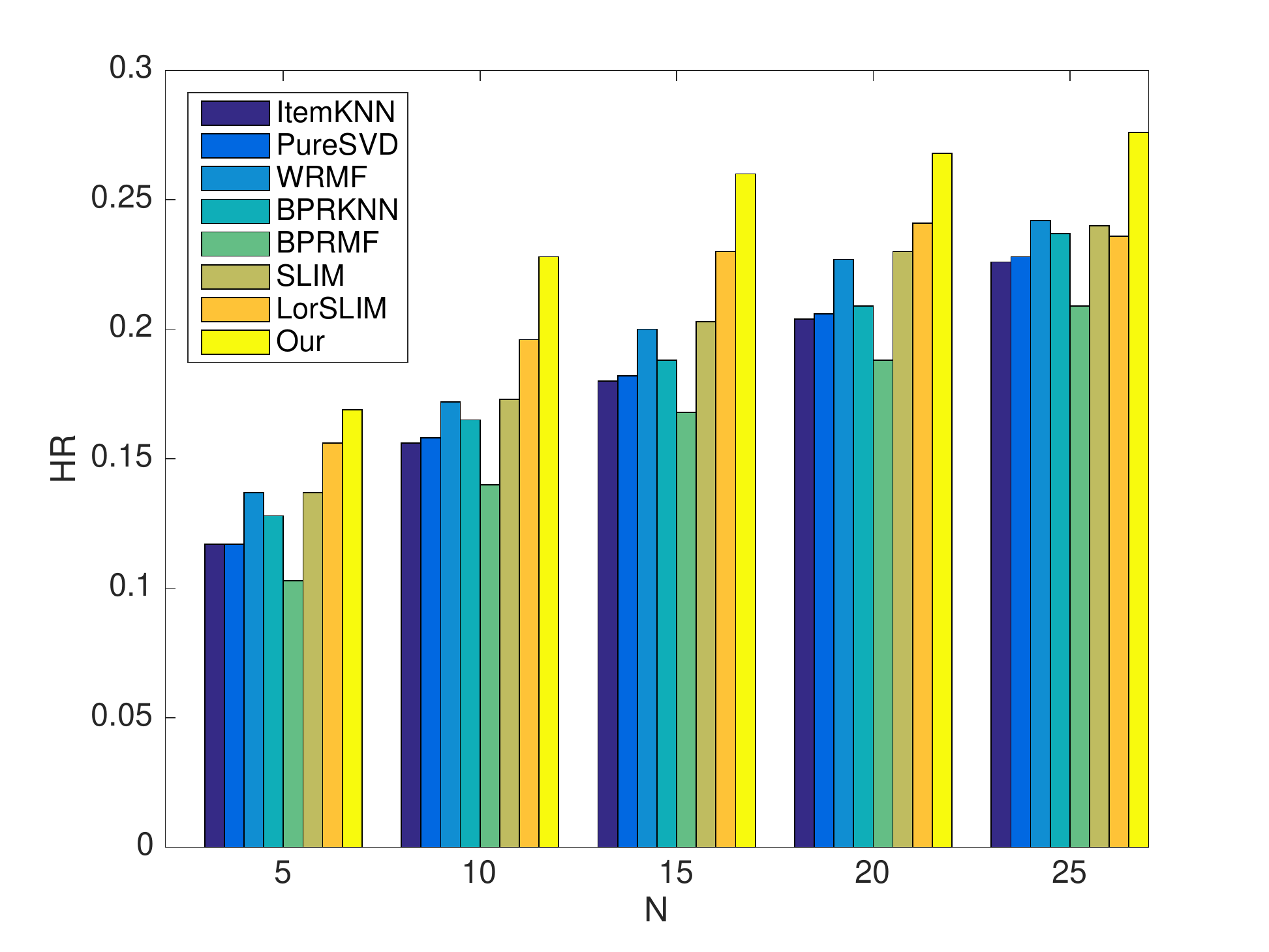}}
\subfigure[Yahoo]{\includegraphics[width=.38\textwidth]{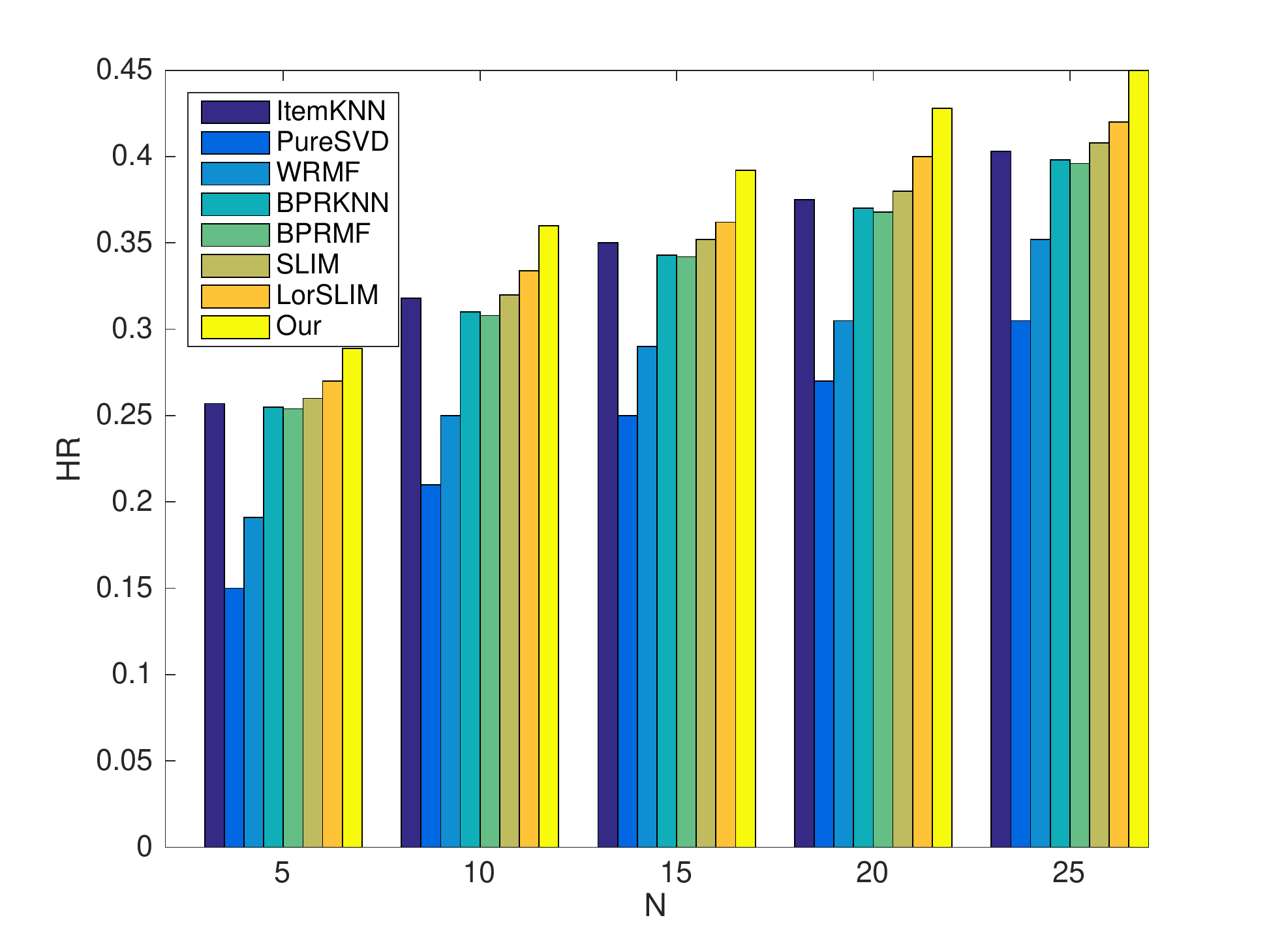}}
\caption{ Performance for Different Values of $N$.}
\label{fig:differentN}
\end{figure*}
\subsection{Comparison Algorithms}
We compare the performance of the proposed method with seven state-of-the-art Top-$N$ recommendation algorithms, including the item neighborhood-based collaborative filtering method ItemKNN \cite{deshpande2004item}, two MF-based methods PureSVD \cite{cremonesi2010performance} and WRMF \cite{hu2008collaborative}, SLIM \cite{ning2011slim} and LorSLIM \cite{cheng2014lorslim}. We also examine two ranking/retrieval criteria based methods BPRMF and BPRKNN \cite{rendle2009bpr}, where Bayesian personalized ranking (BPR) criterion is used which measures the difference between the rankings of user-purchased items and the remaining items.

\section{Results}
\label{discuss}

\subsection{Top-N Recommendation Performance}
We summarize the experimental results of different methods in Table \ref{tab:comp}.
It shows that our algorithm performs the best among all methods across all the datasets\footnote{Codes of our algorithm can be found at https://github.com/sckangz/SDM16}. Specifically, in terms of HR, our method outperforms ItemKNN, PureSVD, WRMF, BPRKNN, BPRMF, SLIM and LorSLIM by 40.41\%, 47.22\%, 34.65\%, 27.99\%, 36.01\%, 25.67\%,  11.66\% on average, respectively, over all the six datasets; with respect to ARHR, the average improvements across all the datasets for ItemKNN, PureSVD, WRMF, BPRKNN, BPRMF, SLIM and LorSLIM are 45.79\%, 56.38\%, 45.43\%, 34.25\%, 46.71\%, 29.41\%, 11.23\%, respectively. This suggests that a closer rank approximation than the nuclear norm is indeed crucial in real applications.

Among seven other algorithms, LorSLIM is a little better than the others. SLIM, BPRMF, and BPRKNN give similar performance. For the three MF-based methods, BPRMF and WMF are better than PureSVD except on lastfm and ML100K. It is interesting to note that the simple itemKNN performs better than BPRMF on Netflix and Yahoo. This could be because in BPRMF , the entire AUC curve is used to measure if the interested items are ranked higher than the rest. However, a good AUC value may not lead to good performance for Top-$N$ recommendation \cite{rendle2009bpr}.

\subsection{Recommendation for Different Top-N}
We show the performance of these algorithms for different values of $N$ (i.e., 5, 10, 15, 20 and 25) on all six datasets in Figure \ref{fig:differentN}. It shows that our algorithm outperforms other methods significantly in all cases. Once again, it demonstrates the importance of good rank approximation.
\subsection{Matrix Reconstruction}
We use ML100K to show how LorSLIM and our method reconstruct the user-item matrix. The density of ML100K is 6.30\% and the mean for those non-zero elements is 3.53. The reconstructed matrix $\hat{X}_{LorSLIM}$ from LorSLIM has a density of 13.61\%, whose non-zero values have a mean of 0.046. For those 6.30\% non-zero entries in $X$, $\hat{X}_{LorSLIM}$ recovers 70.68\% of them and their mean value is 0.0665. In contrast, our proposed algorithm recovers all zero values. The mean of our reconstructed matrix is 0.236. For those 6.30\% non-zero entries in $X$, it gives a mean of 1.338. These facts suggest that our method better recovers $X$ than LorSLIM can do. In other words, LorSLIM loses too much information. This appears to explain the superior performance of our proposed method.  

\subsection{Parameter Effects}
\begin{figure}[!ht]
\centering
\includegraphics[width=.38\textwidth]{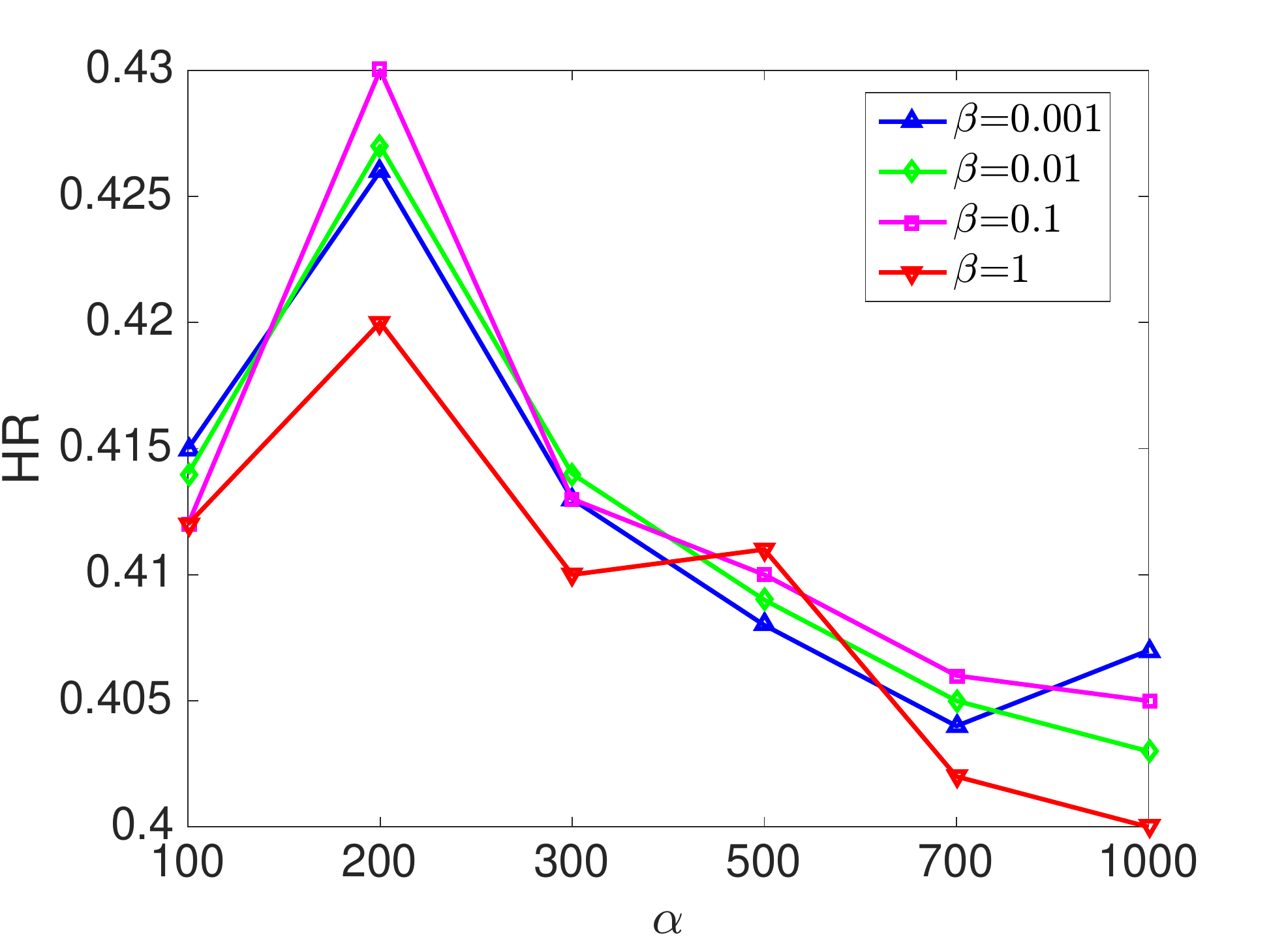}
\caption{ Influence of $\alpha$ and $\beta$ on HR for ML100K dataset.}
\label{parameter}
\end{figure}
Our model involves parameters $\alpha$, $\beta$. We also introduce an auxiliary parameter $\mu$ in ALM algorithm. Some previous studies have pointed out that a dynamical $\mu$ is preferred in practice. Hence we increase $\mu$ at a rate of $\gamma$ with a value 1.1, which is a popular choice in the literature. For each possible combination of $\alpha$, $\beta$, we can use grid search to find the optimal initial value $\mu^0$. 

In Figure \ref{parameter}, we depict the effects of different $\alpha$, $\beta$ on dataset ML100K. As can be seen from it, our algorithm performs well over a large range of $\alpha$ and $\beta$. Compared to $\beta$, the result is more sensitive to $\alpha$. The performance keeps increasing as $\alpha$ increase when it is small, then decreases as it become larger. This is because the $l_1$-norm parameter $\alpha$ controls the sparsity of the aggregating matrix. If $\alpha$ is too large, the matrix will be too sparse that nearly no item will be recommended since the coefficients with the target item are all zero. 
\begin{figure}[!ht]
\centering
\includegraphics[width=.38\textwidth]{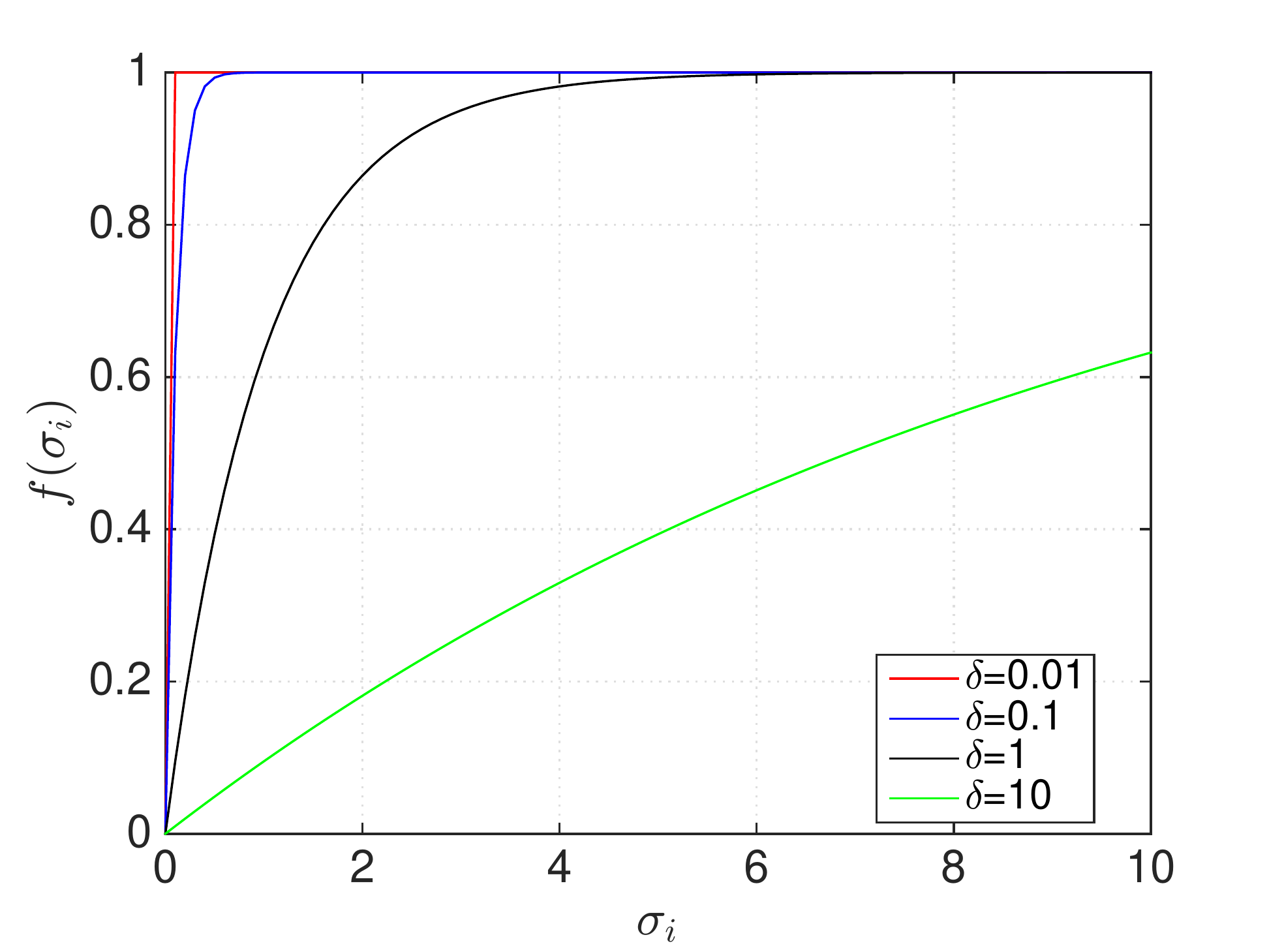}
\caption{Effect of $\delta$ on rank approximation.}
\label{delta}
\end{figure}

Another important parameter is $\delta$ in our rank approximation, which measures how close of our rank relaxation to the true rank. Generally speaking, it is always safe to choose a small value, although $\delta$ can be big if the singular values are big or the size of matrix is big. If $\delta$ is too small, it may incur some numerical issues. Figure \ref{delta} displays the influence of $\delta$ on the rank approximation. It can be seen that $f$ can match the rank function closely when $\delta\leq0.1$. For our previous experimental results, $\delta=0.1$ is applied, which results in an approximation error of $0.05$.

\section{Conclusion}
\label{conclude}   
In this paper, we propose a novel rank relaxation to solve the Top-$N$ recommendation problem. This approximation addresses the limitations of the nuclear norm by mimicing the behavior of the true rank function. We show empirically that this nonconvex rank approximation can substantially improve the quality of Top-$N$ recommendation. This surrogate for the rank function of a matrix may as well benefit a number of other problems, such as robust PCA and robust subspace clustering.  
\section{Acknowledgments}
This work is supported by the U.S. National Science Foundation under Grant IIS 1218712.

%
%

\bibliographystyle{IEEEtran}
\bibliography{recom}
\end{document}